# Waveguide taper engineering using coordinate transformation technology

**Paul-Henri Tichit, Shah Nawaz Burokur, and André de Lustrac**

*Institut d'Electronique Fondamentale, Centre Scientifique d'Orsay, 91405 Orsay cedex, France*
*paul-henri.tichit@u-psud.fr, sburokur@u-paris10.fr, andre.de-lustrac@u-psud.fr*

**Abstract:** Spatial coordinate transformation is a suitable tool for the design of complex electromagnetic structures. In this paper, we define three spatial coordinate transformations which show the possibility of designing a taper between two different waveguides. A parametric study is presented for the three transformations and we propose achievable values of permittivity and permeability that can be obtained with existing metamaterials. The performances of such defined structures are demonstrated by finite element numerical simulations.

## 1. Introduction

Although now well-known for 30 years, the invariance of Maxwell's equation has been redesigned since 2006 by J. B. Pendry *et al*. [1] and U. Leonhardt [2]. Some proposals of devices using this method have been presented for cloaking [1, 3, 4], concentrators [5], beam bends and expanders [6-10], rotators [11], channels [12] and ultra-directive antennas [13-14]. Thus transformation optics appears to be a convenient tool to design devices or components with special properties difficult to obtain with naturally occurring materials.

Theoretically, the coordinate transformation method consists in generating a new transformed space from an initial one where solutions of Maxwell's equations are known. First of all, we imagine a virtual space with the desired topological properties. The main goal is to create a new space which contains the underlying physics. We gather all these properties in the metric tensor in order to perform the calculations. Although this approach and associated calculations are now well known, Pendry *et al*. [1] have proposed a new interpretation where the new components tensors of permeability and permittivity can be viewed as a material in the original space. It is as if the new material mimics the defined topological space. These mathematical tools have been intensively used since two years for the design of optical devices and components [5-14].

In this paper, following the transformation optics approach, we design a taper between two waveguides of different cross sections. Three different transformation techniques are presented so as to achieve a taper between the two waveguides. The media obtained from these three methods presents complex anisotropic permittivity and permeability. However, we show that using an exponential transformation leads to the design of a taper from a material with physically achievable material parameters.

## 2. Transformation formulations

Three different formulations are proposed below to achieve a low reflection taper between two waveguides of different cross sections. Each waveguide is represented by black lines in its respective space given in Cartesian coordinates as depicted in Fig. 1. The aim is to connect each horizontal lines of each space to assume transmission of electromagnetic waves. Thus in geometric approximation, each ray of light in the first waveguide is guided into the second one by green lines representing the taper. For the first formulation, we assume a linear transformation by connecting the two spaces with straight lines as shown in Fig. 1(a). The second formulation uses a parabolic transformation to achieve the connection (Fig. 1(b)). For the third one, an exponential transformation is defined as shown in Fig. 1(c). In all three cases, the geometrical properties of the schema under analysis remain unchanged. The width of the input and output waveguides is respectively noted *a* and *b*, and the length of the taper in all three cases is taken to be *l*. Mathematical expressions defining each formulation of the transformation approaches are given in Fig. 1. *x'*, *y'* and *z'* are the coordinates in the transformed (new) space and *x*, *y* and *z* are those in the initial space. As it can be observed from the mathematical expressions, the different formulations depend on the geometric parameters (*a*, *b*, *l*).

Each transformation leads to a material with specific properties that can play the role of the desired taper. The transformation approach can be summarized in two main points. First of all we determine the Jacobian matrix of each transformation formulation so as to obtain the properties of the "taper space". Thus we can use the transformation expressed in (1) to obtain the permittivity and permeability tensors in the initial space which are *x* and *y* dependent.

$$\varepsilon^{i'j'} = \frac{J_i^{i'} J_j^{j'} \varepsilon_0 \delta^{ij}}{\det(J)} \quad \text{and} \quad \mu^{i'j'} = \frac{J_i^{i'} J_j^{j'} \mu_0 \delta^{ij}}{\det(J)} \quad \text{with} \quad J_\alpha^{\alpha'} = \frac{\partial x'^\alpha}{\partial x^\alpha} \quad (1)$$

where $J_\alpha^{\alpha'}$ and $\delta^{ij}$ are, respectively, the Jacobian transformation matrix of the expressions given in Fig. 1 and the Kronecker symbol.

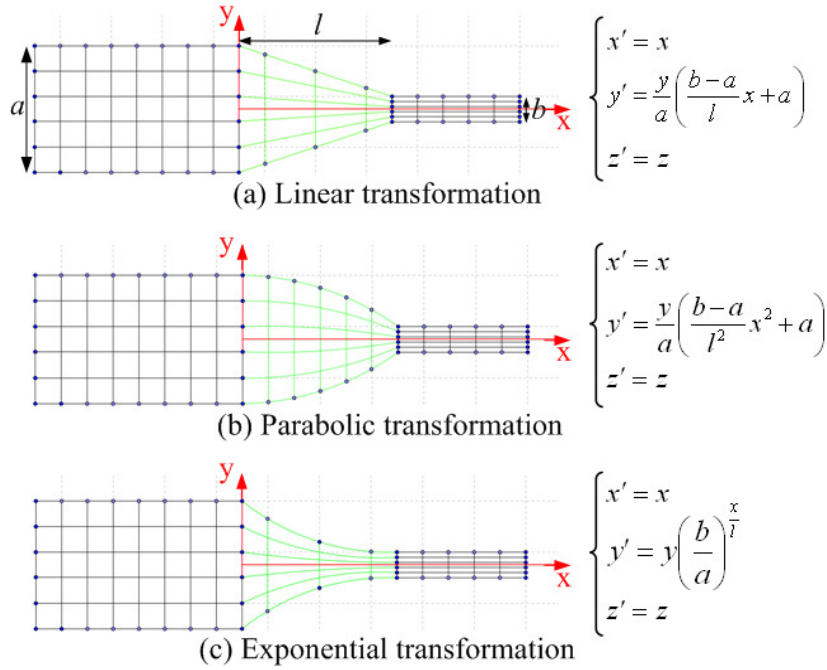

Fig. 1. Transformed taper (green lines) between two waveguides (black lines) with different cross sections. (a) Linear, (b) parabolic, and (c) exponential transformation formulation.

The second step consists in calculating the new permittivity and permeability tensors in the coordinate system (*x'*, *y'*) so as to mimic the transformed space. At this stage, we have designed a material with specific desired physical properties. This material can then be described by a permeability and permittivity tensors $\overline{\overline{\varepsilon}} = \overline{\overline{\theta}} \varepsilon_0$ and $\overline{\overline{\mu}} = \overline{\overline{\theta}} \mu_0$. In order to simplify the different calculations, we take $\dfrac{J_i^{i'} J_j^{j'} \delta^{ij}}{\det(J)} = \theta^{i'j'}$ with

$$\bar{\bar{\theta}} = \begin{pmatrix} \theta_{xx}(x') & \theta_{xy}(x',y') & 0 \\ \theta_{xy}(x',y') & \theta_{yy}(x',y') & 0 \\ 0 & 0 & \theta_{zz}(x') \end{pmatrix} \quad (2)$$

The components values of $\bar{\bar{\theta}}$ tensor are given in Table 1 where a nondiagonal term ($\theta_{xy}$) appears. This nondiagonal term is necessary to guide electromagnetic waves in the *x-y* plane like it is the case for this taper.

Table 1. Components values of $\bar{\bar{\theta}}$ tensor for the three transformations.

|  | $\theta_{xx}(x') = \theta_{zz}(x')$ | $\theta_{xy}(x',y')$ | $\theta_{yy}(x',y')$ |
|---|---|---|---|
| *Linear transformation* | $\dfrac{al}{a(l-x')+bx'}$ | $\theta_{xx}^2 \dfrac{b-a}{al} y'$ | $\dfrac{1}{\theta_{xx}} + \dfrac{\theta_{xy}^2}{\theta_{xx}}$ |
| *Parabolic transformation* | $\dfrac{al^2}{(al^2 - ax'^2 + bx'^2)}$ | $\dfrac{2\theta_{xx}^2 (b-a) x' y'}{al^2}$ | $\dfrac{1}{\theta_{xx}} + \dfrac{\theta_{xy}^2}{\theta_{xx}}$ |
| *Exponential transformation* | $\left(\dfrac{b}{a}\right)^{\frac{x'}{l}}$ | $\dfrac{\theta_{xx} y' \ln(\dfrac{b}{a})}{l}$ | $\dfrac{1}{\theta_{xx}} + \dfrac{\theta_{xy}^2}{\theta_{xx}}$ |

## 3. Simulations and results

To verify the results expressed in the previous section, finite element method based commercial software Comsol MULTIPHYSICS is used to design the described waveguide taper. Simulations are performed in two-dimensional mode for the validation of the proposed material parameters. Port boundaries are used to excite the first and third Transverse Electric (TE$_1$ and TE$_3$) modes of the input waveguide with the E-field directed along the *z*-axis to verify the conservation of modes through the taper. The waveguides boundaries are assumed as Perfect Electric Conductors (PECs) and matched boundaries conditions are applied to the taper. Verifications are done in the microwave domain for a possible future physical prototyping based on the use of metamaterials [15-20], such as Split Ring Resonators (SRRs), Electric LC resonators (ELCs), cut wire pairs and continuous wires. SRRs [15] are known to produce a magnetic resonance where the permeability ranges from negative to positive values and continuous wires [16] have been presented to exhibit a Drude-like permittivity response with negative values below the plasma frequency. Symmetric cut wire pairs presented in [17] produce also a magnetic resonance where as ELCs [18] produce an electric resonance. Asymmetric cut wire pairs recently proposed in [19-20] have experimentally demonstrated both electric and magnetic resonances.

The waveguides widths are chosen to be a = 10 cm and b = 2 cm with respectively 1.5 GHz and 7.5 GHz cutoff frequencies. Length of the taper is chosen as *l* = 5 cm allowing to generate the entire spatial dependence of the material parameters $\theta_{xx}(x')$, $\theta_{zz}(x')$, $\theta_{xy}(x',y')$ and $\theta_{yy}(x',y')$ as shown in Fig. 2. These distributions are plotted from the expressions given in Table 1. Values of permittivities and permeabilities presented in Fig. 2 account for the control of the electromagnetic field in the taper and the conservation of the propagating modes from waveguide 1 to waveguide 2. Although same spatial distribution profile can be observed for the three different formulations, parameters values are completely different. For linear and parabolic transformations, values of $\mu_{yy}$ are too high to be physically achievable with existing metamaterials. However, it is clear that the exponential transformation leads to values more easily achievable with metamaterials. Moreover, the physical realization of such a metamaterial taper will be facilitated by the slow variation of the material parameters, implying a gradual variation of the geometrical parameters of metamaterial inclusions. We shall note that the components are calculated in the Cartesian system and obey the following dispersion relation in the transverse electric (TE) mode:

$$\varepsilon_{zz}\left(\mu_{xy}^2 - \mu_{xx}\mu_{yy}\right) + \mu_{xx}k_x^2 + k_y\left(\mu_{yy}k_y - 2\mu_{xy}k_x\right) = 0 \qquad (3)$$

This equation is obtained from the propagation equation and describes the control of electromagnetic waves in the material. This relation is also important for a future reduction of parameters, which can be done by simplifying the nondiagonal parameter $\theta_{xy}$ to a closed interval near zero when choosing appropriately the length of the taper. For example, by bounding the nondiagonal term of the exponential formulation in Table 1, the condition

$l > \dfrac{a^2}{b}\dfrac{\left|\ln\left(\dfrac{b}{a}\right)\right|}{2\Delta}$ leads to $-\Delta < \mu_{xy} < \Delta$ where $\Delta$ can be very close to zero.

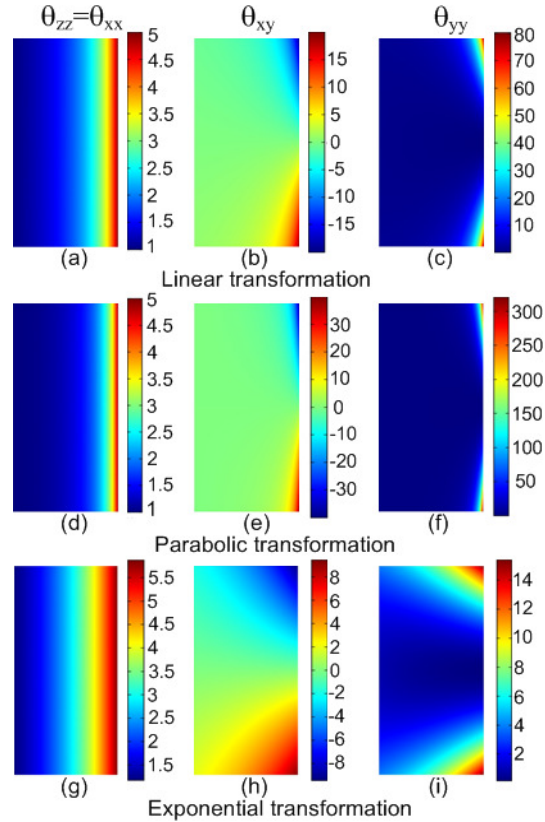

Fig. 2. Components of the permittivity and permeability tensors $\bar{\bar{\theta}}$ for the three transformations with $a = 10$ cm, $b = 2$ cm and $l = 5$ cm.

Simulation results of the E-field distribution of the structure under study for the three transformations are presented in Fig. 3. The distributions in the tapered waveguides are compared to a non tapered case at 10 GHz and 30 GHz for the fundamental ($TE_1$) excitation mode and at 30 GHz for the third ($TE_3$) excitation mode. Concerning the non tapered junction waveguides, we can observe phase distortions caused by reflections at the junction from the bigger waveguide to the smaller one (Figs. 3(a)-(c)). These distortions become more severe at higher frequencies (30 GHz). However, simulations performed on the tapered waveguides (Figs. 3(d)-(l)) illustrate that electromagnetic waves are properly guided from one waveguide to the other without any impact on the guided mode when the transformed medium is embedded between the two waveguides. The difference in the transformation formulations indicates a change in the path of electromagnetic waves in the tapered section, highlighted by the shaded gray area in Fig. 3. Increasing the frequency improve the transmission between the two waveguides through the tapered section. This phenomenon can be observed when we compare the E-field distributions at 10 GHz and 30 GHz. At 10 GHz, a slight impedance mismatch between the taper output and the small waveguide input can be observed. This phenomenon decreases at higher frequencies, as illustrated for 30 GHz. Fig. 3 shows the efficiency of the material parameters defined by coordinate transformation technology.

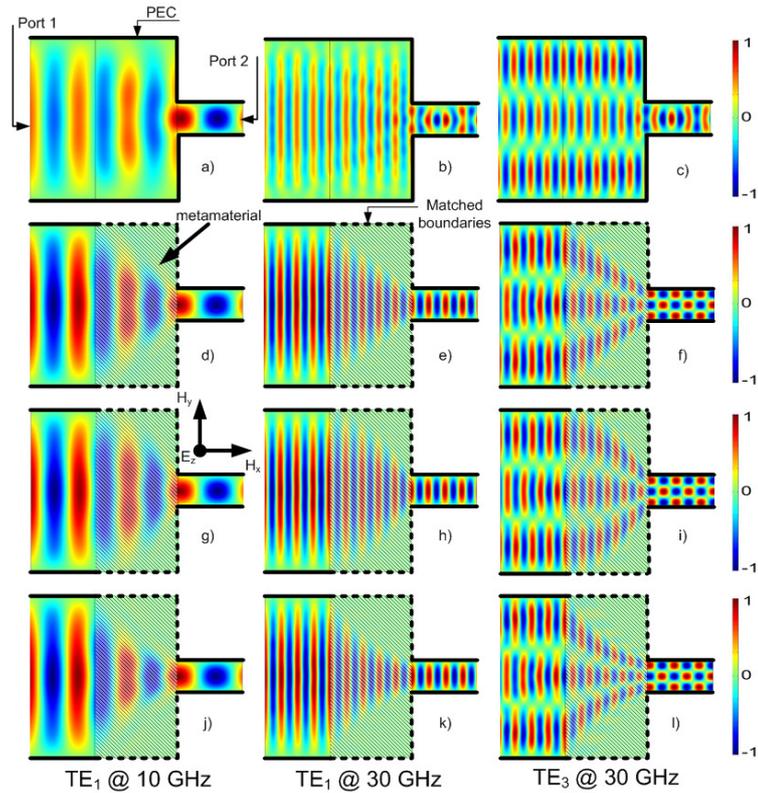

Fig. 3. Normalized E-field distribution for TE polarization. (a)-(c) Non tapered junction waveguides, (d)-(f) Tapered junction waveguides with linear transformation. (g)-(i) Tapered junction waveguides with parabolic transformation. (j)-(l) Tapered junction waveguides with exponential transformation.

## 4. Conclusion

Spatial coordinate transformation is applied to design a taper between two waveguides with different cross sections. Three transformation formulations have been tested; a linear, a parabolic and an exponential one. Each formulation is used to generate material parameters and it has been observed that the parameters profile depend on the formulation considered. Best tradeoff is obtained with the exponential formulation with a slow variation of the material parameters. This slow variation is very interesting if we consider physical realization of such a structure using metamaterials as SRRs, ELCs, cut and continuous wires. Numerical simulations using the material parameters have been performed to show how electromagnetic waves are guided between two waveguides through the taper. Compared to an abrupt (non tapered) junction where phase distortions appear, transformation optics help to guide properly electromagnetic waves from one waveguide to the other without any impact on the guided propagating modes.